# Effect of strength of gravitational field on the rate of chemical reactions

Mirza Wasif Baig*

E-mail: wasifahmed797@gmail.com

**Abstract:** The magnitude of the rate of chemical reactions also depends on the position in the gravitational field where a chemical reaction is being carried out. The rate of chemical reaction conducted at a stronger gravitational field, i.e., near the surface of some heavy planet, is slower than the rate of reaction conducted at a weaker gravitational field, i.e., away from the surface of a heavy plant, provided temperature and pressure are kept constant at two positions in the gravitational field. The effect of gravity on the rates of reactions has been shown by formulating the rate constants from almost all types of reaction rate theories, i.e., transition state theory, collision theory, Rice-Ramsperger-Kassel-Marcus (RRKM), and Marcus's theory, in the language of the general theory of relativity. The gravitational transformation of the Boltzmann constant and the energy quantum levels of molecules have been developed quantum mechanically. A gravitational transformation of thermodynamic state functions has been formulated that successfully explains the quasi-equilibrium existing between reactants and the activated complex at different gravitational fields. Gravitational mass dilation has been developed that explains that at weaker gravitational fields, the transition states possess more kinetic energy to sweep translation on the reaction coordinate, resulting in the faster conversion of reactants into products. The gravitational transformation of the half-life equation shows gravitational time dilation for the half-life period of chemical reactions and thus renders the general theory of relativity and the present theory in accord with each other.

## 1 Introduction

The special theory of relativity proposed by Einstein appeared in 1905 [1, 2] and discarded the absolute notions of space and time. [3] General relativity was born because of efforts to extend the special theory of relativity to non-inertial frames. [4] It describes gravity as an effect rather than a force. It completely interweaves space and time into one entity: space-time. This space-time behaves as a flexible fabric. Warps and curves in this fabric of space-time are the origin of gravity. The geometry of this four-dimensional space-time fabric completely defines the behavior of gravity. General relativity explains that clocks located near the surface of a heavy planet, i.e., those with stronger gravitational potential, run slower than clocks located at a certain appreciable height from the surface of a planet, i.e., those with weaker gravitational potential. Gravitational time dilation is a kind of time dilation that describes an actual difference in elapsed time between two events as measured by observers situated at varying distances from a gravitating mass. The closer the clock is to the source of gravity i.e., at a stronger gravitational potential, the slower the time passes for it. Moving the same clock away from the source of gravitation i.e., at a weaker gravitational potential, speeds up its time flow [3]. This effect has also already been experimentally verified [5]. In the past, there have been few attempts to study the effect of gravity on chemical reactions [6, 7], but a theory that can truly explain how gravity will really affect the rate of chemical reactions is missing so far. More than three decades before, Ohsumi also made an attempt to explain chemical reactions from the point of special and general relativity [8], but without taking into account relativistic and gravitational time dilation effects. Recently, one attempt has been made to study the effect of gravity on chemical reactions, but the claim of that study totally disagrees with that of general relativity [9]. Since general relativity explains that time slows down near the surface of heavy planets where gravity is strong, chemical processes should slow down near the source of gravitation, i.e., at a stronger gravitational field [3]. But P. Lecca has claimed that intense gravity will speed up rates of chemical reactions [9], which is opposite to the gravitational time dilation phenomenon [3]. So, the present study is the first attempt to invoke gravitational time dilation in chemical kinetics that can explain the effect of gravity on rates of chemical reactions in the gravitational potential of an arbitrary massive planet with mass $M$.

## 2 Theory

To compare the rates of chemical reactions at different positions in the gravitational field where gravity is stronger at positions closer to the surface of a heavy planet and weaker at appreciable heights from the surface of a planet, temperature and pressure at two positions should be kept the same, as rates of chemical reactions are affected by variations in temperature and pressure [10]. Since gravity is defined as an effect that emerges as an entropic force whose origin is explained by employing a canonical ensemble, thus ensuring temperature to remain constant in the entire gravitational field [11], Black hole thermodynamics [12, 13] also supports the fact that temperature is not affected by the strength of the gravitational field. This is quite similar to the Lorentz invariance of temperature [14], which has been proven to be Lorentz invariant using quantum statistical treatment centered on the zeroth law of thermodynamics [15] and phenomenological analysis [16]. The relativistic theory of chemical kinetics, which is the first attempt to explain relativistic time dilation in chemical and nuclear reactions [17], also fully supports Lorentz invariance of temperature. To explain gravitational time dilation, we first need to consider the relativistic increase in mass for observers moving at fractions of the speed of light. For a particle moving at a fraction of the speed of light, an increase in its mass will be given, which is defined as:

$$m_u = 1/\sqrt{1 - u^2/c^2}\, m_0 \quad (1)$$

When particles are moving at a sufficient, weaker speed, there is an increase in their mass. This relativistic mass for particles moving at weaker speeds is a consequence of the special theory of relativity [1-3]. To have compatibility between the special theory of relativity and general relativity, mass dilation for particles in a gravitational field should exist, like gravitational time dilation [3]. It can be developed considering a simple gedanken experiment. It is known that when a beam of photons is moved straight from a region of stronger gravitational potential (near the source of gravity) to a region of weaker gravitational potential (away from the source of gravity), it goes under redshift, i.e., their wavelength decreases at a weaker gravitational potential [18]. The gravitational attraction that leads to a general mass accretion of massive particles has already been reported [19]. Now consider a thought experiment in which either a beam of electrons or atoms is projected against gravitational potential (from a region of stronger gravitational potential to a region of weaker gravitational potential) due to a gravitational red shift. The de Broglie wavelength of either atoms or electrons will be greater at weaker gravitational potential than at stronger gravitational potential. This will result in an increase in the gravitational mass of electrons and atoms when they are accelerated towards stronger gravitational potential due to the de Broglie relation [20]. This leads to the mathematical formulation of gravitational mass dilation as follows:

$$m_s = 1/\sqrt{1 - 2GM/rc^2}\ m_h \quad (2)$$

To explain the effect of gravity on chemical reactions, a gedanken experiment is carried out in which a chemical system is drifted towards the source of gravity of some heavy planet with mass $M$. Or, in other words, one chemical reaction is carried out at a height "$h$" from the surface of a heavy planet, i.e., at a weaker gravitational field, and another chemical reaction is carried out at the surface of a heavy planet with mass $M$, provided temperature and pressure are the same at two positions in the gravitational field. All physical quantities of this chemical system that undergo a change in magnitude with a change in the strength of the gravitational field will be defined by a subscript "$s$" at a stronger gravitational field, i.e., near the surface of a heavy planet, and denoted by a subscript "$h$" at a weaker gravitational field, i.e., away from the surface of a heavy planet. In this work all mathematical relations with subscripts "$s$" and "$h$" refer to the position of the chemical system at the surface of a heavy planet and at height "$h$" from the surface of a heavy planet, respectively. Gravitational transformations express how any particular physical quantity "$Y$" of the chemical system under study changes in magnitude when moving the chemical system from a weaker gravitational potential "$Y_h$" to a stronger gravitational potential "$Y_s$". In gravitational transformations "$Y_s$" and "$Y_h$" are related to one another by a factor "$\xi = 1/\sqrt{1 - 2GM/rc^2}$" where $M$ is the mass of the heavy planet and $r$ is the radial coordinate of the observer [3]. Chemical reactions in this thought experiment are conducted in the gravitational field of a planet with a mass $M$. When some physical quantity "$Y$" of the chemical system under study increases in magnitude on moving towards a gravity source, i.e., at a stronger gravitational field than gravitational transformation, the physical quantity "$Y$" of the chemical system will be mathematically expressed as "$Y_s = \xi\, Y_h$". Like in the gravitational transformation of mass given by Eq. (2), the mass of a particle "$m_s$" at a stronger gravitational field is greater than the mass of a particle "$m_h$" at a weaker gravitational field because "$m_h$" is multiplied by a mathematical factor "$\xi = 1/\sqrt{1 - 2GM/rc^2}$" that will always be greater than unity. Similarly, in other kinds of gravitational transformation, when some physical quantity "$Y$" of the chemical system under study decreases in magnitude on moving towards the gravity source, i.e., at a stronger gravitational field, the physical quantity "$Y$" of the chemical system will be mathematically expressed as $Y_s = \xi^{-1}\, Y_h$. So, in this kind of gravitational transformation, "$Y_s$" will be smaller in magnitude than "$Y_h$" because later is multiplied by a mathematical factor "$\xi^{-1} = \sqrt{1 - 2GM/rc^2}$" that will always be smaller than unity. For convenience, let's call "$Y_s = \xi\, Y_h$" $\boldsymbol{\alpha}$-gravitational transformation and "$Y_s = \xi^{-1}\, Y_h$" β-gravitational transformation. So, the gravitational transformation of mass is an $\boldsymbol{\alpha}$-gravitational transformation. The exact value of any particular physical quantity "$Y$" of the chemical system will be one in space in the absence of any gravitational field.

## 3 Gravitational transformations in Statistical Mechanics

Heisenberg time energy uncertainty principle dictates that greater the spacing between two quantum levels of a system, longer the system can survive in that excited quantum level. Electronic transitions are quickest most quantum jumps in molecules followed by vibrational transitions and at the end by rotational transitions. This hierarchy of quantum jumps in molecules is a direct consequence of uncertainty principle [21-23]. If a molecule in an excited $j\text{–}th$ quantum state is moved towards near the source of gravity i.e. at stronger gravitational potential from some weaker gravitational potential with appreciable height "$h$" then lifetime of this excited state in $j\text{–}th$ quantum level will be stretched due to slower passage of time near the surface of planet i.e. stronger gravitational field and will be mathematically expressed as;

$$\langle \Delta t \rangle_s = \xi \langle \Delta t \rangle_h \tag{3}$$

For this $j\text{–}th$ quantum state Heisenberg time energy uncertainty relation can be formulated as [24],

$$\langle \Delta \varepsilon_j \rangle_s \langle \Delta t \rangle_s \geq \hbar \tag{4}$$

Since gravitational transformation of lifetime of excited particle in some $j\text{–}th$ quantum state has $\boldsymbol{\alpha}$-gravitational transformation i.e. "$\langle \Delta t \rangle_s = \xi \langle \Delta t \rangle_h$" [3]. In order to keep Heisenberg time energy uncertainty valid in entire gravitational field of planet with mass $M$ energy spacing between quantum will have β-gravitational transformation of i.e.

$$\langle \Delta \varepsilon_j \rangle_s = \xi^{-1} \langle \Delta \varepsilon_j \rangle_h \tag{5}$$

Thus from Eq. (5) it can be inferred that at near the source of gravity i.e. at stronger gravitational field energy spacing between the quantum states will decrease resulting in slower decay of higher excited quantum states to lower quantum states. From spectroscopic signatures of molecules following inequalities exists in nature for electronic, vibrational and rotational transitions of molecules at room temperature which can be written for stronger gravitational potential as [24]

$$(\Delta\varepsilon_{elec})_s >>> (k_B)_s T \tag{6}$$
$$(\Delta\varepsilon_{vib})_s > (k_B)_s T \tag{7}$$
$$(\Delta\varepsilon_{rot})_s \cong (k_B)_s T \tag{8}$$

Inequalities of Eq. s. (6), (7) and (8) for electronic, vibrational and rotational transitions of molecules at room temperature should symmetrically also exist at weaker gravitational field i.e.,

$$(\Delta\varepsilon_{elec})_h >>> (k_B)_h T \tag{9}$$
$$(\Delta\varepsilon_{vib})_h > (k_B)_h T \tag{10}$$
$$(\Delta\varepsilon_{rot})_h \cong (k_B)_h T \tag{11}$$

β-gravitational transformation of energy spacing will result in β-gravitational transformation of Boltzmann constant i.e. $\langle k_B \rangle_s = \xi^{-1} \langle k_B \rangle_h$. Thus, moving the system towards the source of gravity i.e. stronger gravitational potential also results in decrease in magnitude of Boltzmann constant along with the squeezing of energy spacing of quantum states.

### 3.1 Maxwell Boltzmann Distribution Law

Maxwell Boltzmann Distribution law that mathematically defines population of particles in $j\text{–}th$ quantum state with energy $\varepsilon_j$ at stronger gravitational potential can be defined as [25],

$$\langle n_j \rangle_s = exp\left(-\frac{(\varepsilon_j)_s}{(k_B)_s T}\right) \tag{12}$$

Equating β-gravitational transformations of energy spacing between different quantum levels of molecules and Boltzmann constant in Eq. (12) mathematically shows that gravitational field will not affect the population of molecules or particles in any $j\text{–}th$ quantum level i.e.

$$\langle n_j \rangle_s = \langle n_j \rangle_h = n_j \tag{13}$$

### 3.2 Molecular Partition Function

Maxwell Boltzmann statistics is most successful theoretical explanation of distribution of atoms and molecules among various energy states accessible to them in thermal equilibrium. Weaker temperature and low density switch off the quantum effects [25-26]. Product of translational, rotational, vibrational, and electronic partition functions gives total partition function for Maxwell Boltzmann Statistics. When a chemical system is drifted towards stronger gravitational potential then total partition function can be mathematically expressed in terms of individual translational, rotational, vibrational and electronic partition functions as,

$$(Q_{total})_l = \frac{1}{N!}(q^T)_s^N (q^R)_s^N (q^V)_s^N (q^E)_s^N \tag{14}$$

### 3.2.1 Translational Partition Function

When system of molecules of mass $m$ interacting in volume $V$ is drifted towards stronger gravitational field the translational partition function for them can be defined as, [25-26]

$$(q^T)_s = \left(\frac{2\pi m_s (k_B)_s T}{h^2}\right)^{\frac{3}{2}} V \tag{15}$$

Equating $\boldsymbol{\alpha}$-gravitational transformations of mass and β-gravitational transformation of Boltzmann constant in Eq. (15) mathematically shows that gravitational field will not affect the translation molecular partition of molecules. i.e.

$$(q^T)_s = (q^T)_h = q^T \tag{16}$$

### 3.2.2 Rotational temperature and Rotational Partition Function

When system of rotating diatomic molecules is drifted towards source of gravity at then at stronger gravitational potential their rotational temperature can be formulated as, [25-26]

$$(\Theta_R)_s = \frac{hcB_s}{(k_B)_s} \tag{17}$$

While rotational constant at stronger gravitational potential can be formulated as,

$$B_s = \frac{h}{8\pi^2 c I_s} \tag{18}$$

Since due to $\boldsymbol{\alpha}$-gravitational transformation of mass of electron mass at stronger gravitational field will increase so according to Heisenberg's Uncertainty principle i.e., "$m\Delta v\Delta x \geq \hbar/2$" velocity of electron revolving should decrease with same factor "$\xi^{-1}$" rendering the total momentum of electron unchanged that keeps bond length same at all gravitational potentials. $\boldsymbol{\alpha}$-gravitational transformations of mass will also make moment of inertia to have $\boldsymbol{\alpha}$-gravitational transformation i.e. $I_s = \xi I_h$ [25-26]. $\boldsymbol{\alpha}$-gravitational transformation of moment of inertia will make rotational constant "$B_s$" to have β-gravitational transformation i.e.

$$B_s = \xi^{-1} B_h \tag{19}$$

Equating β-gravitational transformations of rotational constant and of Boltzmann constant in Eq. (17) mathematically shows that gravitational field will not affect the rotational temperature of molecules. i.e.

$$(\Theta_R)_s = (\Theta_R)_h = \Theta_R \tag{20}$$

Rotational partition function in terms of rotational temperature at stronger gravitational potential can be defined as,

$$(q^R)_s = \left(\frac{T}{\sigma(\Theta_R)_s}\right) \tag{21}$$

As from Eq. (21) it follows that rotational will remains same in all gravitational potential irrespective of its position in it i.e.,

$$(q^R)_s = (q^R)_h = q^R \quad (22)$$

### 3.2.3 Vibrational temperature and Vibrational Rotational Partition Function

When system containing vibrating molecules is drifted towards source of gravity i.e. near to the surface of planet then vibrational partition function at stronger gravitational potential will be mathematically formulated as, [25-26]

$$(q^V)_s = \frac{1}{1 - \frac{(\Theta_{vib})_s}{T}} \quad (23)$$

Vibrational temperature for these system of vibrating molecules at stronger gravitational potential will be formulated as;

$$(\Theta_{vib})_s = \frac{h\upsilon_s}{(k_B)_s} \quad (24)$$

Since due to $\boldsymbol{\alpha}$-gravitational transformation of time frequency being inversely related to time will have β-gravitational transformation $\upsilon_s = \xi^{-1}\upsilon_h$. So, frequency of molecules behaving as harmonic oscillator at stronger gravitational potential will decrease because mass of atoms will increase at stronger gravitational potential, and they will oscillate much slower at stronger gravitational potential than at weaker gravitational potential.
Thus, β-gravitational transformations of Boltzmann constant and gravitational frequency will keep vibrational temperature constant throughout entire gravitational field i.e.,

$$(\Theta_{vib})_s = (\Theta_{vib})_h = \Theta_{vib} \quad (25)$$

As it follows from Eq. (25) that vibrational temperature remains unchanged at different gravitational potentials, so this renders vibrational partition function remain unchanged with change in strength of gravitational field i.e.

$$(q^V)_s = (q^V)_h = q^V \quad (26)$$

### 3.2.4 Electronic temperature and Electronic Partition Function

When a system of atoms and molecules with well-defined electronic levels is drifted towards source of gravity. Then at stronger gravitational field electronic partition function for these atoms and molecules can be formulated in terms of electronic temperature as, [25-26]

$$(q^E)_s = \sum_i^N g_i \, exp\left(-\frac{(\Theta_E)_s}{T}\right) \tag{27}$$

Electronic temperature for these atoms and molecules at stronger gravitational potential will be,

$$(\Theta_E)_s = \frac{h({\varepsilon_i}^E)_s}{(k_B)_s} \tag{28}$$

Again β-gravitational transformations of Boltzmann constant and energy levels straight forward give,

$$(\Theta_E)_s = (\Theta_E)_h = \Theta_E \tag{29}$$

From Eq. (29) it is clear that electronic temperature will not be affected by force of gravity which will render electronic partition function to have same magnitude in entire gravitational field.

$$(q^E)_s = (q^E)_h = q^E \tag{30}$$

From all respective individual partition functions, it can be inferred that total molecular partition function will remain same in the entire gravitational field i.e.,

$$(Q_{total})_s = (Q_{total})_h = Q_{total} = Q \tag{31}$$

Thus, gravity has no effect on magnitude of molecular partition function of chemical system.

## 4 Gravitational transformations in Statistical Thermodynamics

Statistical mechanics gives the molecular-level view of all the macroscopic thermodynamic quantities such as internal energy, free energy and entropy. In statistical thermodynamics all properties of system in thermodynamics equilibrium are encoded in terms of partition function $Q$. All thermodynamic state variables and equilibrium constant is mathematically expressed in terms of partition function [25-26]. For a chemical system drifted towards source of gravity. All thermodynamic state functions i.e. integral energy $U$; enthalpy $H$; entropy $S$; Gibbs Helmholtz energy $A$; and Gibbs free energy $G$ at stronger gravitational potential will be mathematically expressed as,

$$U_s = -N(k_B)_s \left(\frac{\partial Q}{\partial T}\right)_V \tag{32}$$

$$H_s = T\left[(k_B)_s T\left(\frac{\partial}{\partial T} ln\, Q\right)_V + V(k_B)_s\left(\frac{\partial}{\partial V} ln\, Q\right)_T\right] \tag{33}$$

$$S_s = (k_B)_s T\left(\frac{\partial \, ln\, Q}{\partial T}\right) + (k_B)_s \, ln\, Q \tag{34}$$

$$A_s = (k_B)_s T \ ln \ Q \quad (35)$$

$$G_s = (k_B)_s T \left[ (ln \ Q) - V \left( \frac{\partial}{\partial V} ln \ Q \right)_T \right] \quad (36)$$

β-gravitational transformation of Boltzmann constant will straight forward make all thermodynamic state functions to exhibit β-gravitational transformations i.e.

$$U_s = \xi^{-1} U_h \quad (37)$$

$$H_s = \xi^{-1} H_h \quad (38)$$

$$S_s = \xi^{-1} S_h \quad (39)$$

$$A_s = \xi^{-1} A_h \quad (40)$$

$$G_s = \xi^{-1} G_h \quad (41)$$

Thus, all thermodynamic state functions of the chemical system under study will decrease in magnitude when that chemical system will be drifted towards source of gravity i.e. at stronger gravitational field. β-gravitational transformations of these thermodynamic state functions will be used to explain quasi-equilibrium existing between reactants and activated complexes during the course of chemical reaction.

## 5 Gravitational transformations in chemical kinetics

To theoretically explain the effect of the strength of the gravitational field on chemical reactions the necessary mathematical forms of the rate laws in four basic theories of chemical kinetics, meeting the requirements of general relativity, are derived in the following.

### 5.1 Gravitational transformation of rate constant from Transition state theory

Transition state theory which separates reactants and products on potential energy surface while taking in account Born-Oppenheimer approximation formulates an expression for thermal rate constant. Maxwell Boltzmann Distribution is used to distribute reactants, products, and transition states among their different quantum states even in the absence of equilibrium between reactant and product molecules [27-29]. There exists special type of equilibrium between reactants and activated complexes named as quasi-equilibrium. In the transition state motion along the reaction coordinate is separated from the other motions and explicitly treated as translational motion in a classical regime. RRKM theory also exclusively treats the motion of transition state along the reaction coordinate as a simple translational motion [30]. IF a chemical system is drifted towards source of gravity, then according to time energy Heisenberg uncertainty principles at stronger gravitational field lifetime of transition state $(\Delta t)_s$ must be greater than $\hbar/(k_B)_s T$ to execute translation along the reaction coordinate. Mathematically it can be written as $(k_B)_s T (\Delta t)_s \geq \hbar$ which according to gravitational time dilation naturally leads to β-gravitational transformation of Boltzmann constant i.e., $(k_B)_s = \xi^{-1}(k_B)_h$. Eyring equation for $nth$ order reaction in thermodynamic terms, at stronger gravitational field will be defined as [31, 32];

$$(k_n)_s = \kappa \frac{(k_B)_s T}{h[(c^o)]^{(n-1)}} exp\left(\frac{(\triangle S^{\dagger})_s}{R_s}\right) exp\left(\frac{(-\triangle H^{\dagger})_s}{R_s T}\right) \tag{42}$$

Placing β-gravitational transformations of Boltzmann constant, enthalpy and entropy gives,

$$(k_n)_s = \xi^{-1}(k_n)_h \tag{43}$$

Arrhenius factor gives quantitative expression for number of reactant molecules crossing energy barrier and transforming into products. At stronger gravitational potential it transforms as; [24-27]

$$A_s = exp[-(\Delta n - 1)] \frac{(k_B)_s T}{[h(c^o)]^{(n-1)}} exp\left(\frac{(\triangle S^{\dagger})_s}{R_s}\right) \tag{44}$$

Placing β-gravitational transformations the Boltzmann constant, the universal gas constant and entropy gives.

$$A_s = \xi^{-1} A_h \tag{45}$$

### 5.2 Gravitational transformation of rate constant from Collision Theory of Bimolecular Reactions

Consider a chemical system under study in which bimolecular chemical reaction is conducted i.e. A + B ⟶ P. Best canidiate for explaining kinetics of such reactions is collision theory. The collision theory for reaction rates treats the molecules of reactants as hard spheres colliding with each other expressing the rate of chemical reaction with number of collisions. The theory mathematically expresses the rate of reaction in terms of three important parameters (i) collision frequency, (ii) collision cross section and (iii) relative velocity. When this chemical system of biomolecular reaction will be drifted towards source of gravity i.e. at stronger gravitational field then rate constant in terms of collision cross section and relative velocity of colliding molecules at stronger gravitational field can be stated as [31, 32],

$$(k_2)_s = N_A \sigma_{AB} (v_r)_s \tag{46}$$

At stronger gravitational field relative velocity between the colliding atoms and molecules can be mathematically expressed as,

$$\langle v_r \rangle_s = \left(\frac{8(k_B)_s T}{\pi\, \mu_s}\right)^{\frac{1}{2}} \tag{47}$$

Placing β-gravitational transformation of the Boltzmann constant and **α**-gravitational transformation of mass in Eq. (47) gives β-gravitational transformation of relative velocity as.

$$\langle v_r \rangle_s = \xi^{-1} \langle v_r \rangle_h \tag{48}$$

Relative velocity between the colliding atoms and molecules will be slowed at stronger gravitational potential due to β-gravitational transformation of Boltzmann constant and mass. This is totally compatible with slower time flow at stronger gravitational field as described by general relativity. At stronger gravitational field events slows down thus collisions among molecules also slows down near the source of gravity. Placing β-gravitational transformation of relative velocity from Eq. (48) in Eq. (46) again gives the β-gravitational transformation of rate constant,

$$\langle k_2 \rangle_s = \xi^{-1} \langle k_2 \rangle_h \tag{49}$$

Collision frequency for bimolecular is mathematically defined in terms of mole densities $\rho_A \rho_B$, collision cross section and relative velocity [31, 32]. At stronger gravitational field collision frequency can be expressed as

$$(Z_{AB})_s = N_A \sigma_{AB} (v_r)_s \rho_A \rho_B \tag{50}$$

Substituting β-gravitational transformation of relative velocity in Eq. (50) reveals that collision frequency also reduces at stronger gravitational potential,

$$(Z_{AB})_s = \xi^{-1} (Z_{AB})_h \tag{51}$$

β-gravitational transformation of collision frequency again explains that at stronger gravitational field molecules will collide at slower rate. Thus, gravity slows down molecular collisions.

### 5.3 Gravitational transformation of rate constant from Marcus Theory of Electron Transfer

Marcus's theory is most successful theoretical model for electron transfer reactions in general and most importantly for outer sphere electron transfer reactions [33-37]. This model is based on the rearrangement of solvent molecules around the reactant ions to configure it for a favorable electron transfer. There is a solvent arrangement around each reactant ion having Gibb's free energy $G$ as a minimum, change in this solvent structure shoots up its free energy. For successful electron transfer the transition state is attained by reduction in the separation between the two reactant ions and reorganization of the solvent structure around each of them. A reaction coordinate for electron transfer can be regarded as combination of these ion-ion separations and solvent reorganization coordinates. Gibb's free energy of reactants and products versus reaction coordinate behave as a parabolic function. Point of intersection of two parabolic curves corresponding to free energies of reactants and products locates the Transition state. Marcus's theory mathematically encodes the activation energy based in terms of reorganization energy. Consider a chemical system in which this $A^{ZA} + B^{ZB} \longrightarrow A^{ZA + \Delta Z} + B^{ZB - \Delta Z}$ electron transfer reaction is carried is drifted towards source of gravity. Then at stronger gravitational field the Marcus expression for rate constant of this electron transfer reactions can be expressed as [33-37],

$$(k_{AB})_s = (Z_{AB})_s\, exp\left(\frac{-\left(\triangle G^o_{AB}\right)_s+\lambda_s)^2}{4\lambda_s R_s T}\right) \tag{52}$$

$\lambda_s$ is the reorganization energy, it quantitatively explains the energy required to reorganize the system structure from initial to final coordinates, avoiding hopping between any two different electronic states. Like other thermodynamic state functions, reorganization energy should have same β-gravitational transformation. Since reorganization energy is composed of solvational and vibrational $(\lambda_o)_s$ and $(\lambda_i)_s$ components respectively. At stronger gravitational field vibrational reorganization energy $(\lambda_i)_s$ can be formulated in terms of reduced force constant $\left(k_j\right)_s$ of the $j\text{–}th$ normal mode coordinates of reactants $q^r_j$ and products $q^p_j$ as,

$$(\lambda_i)_s = \frac{1}{2}\sum_j \left(k_j\right)_s \left(q^r_j - q^p_j\right)^2 \tag{53}$$

At stronger gravitational field reduced force constant $\left(k_j\right)_s$ of the $j\text{–}th$ normal mode can be defined as $\left(k_j\right)_s = 4\pi^2\omega^2_s\mu_s$. β-gravitational transformation of reduced mass "$\mu_s = \xi\, \mu_h$" and oscillation frequency "$\omega_s = \xi^{-1}\omega_h$" will straight forward give β-gravitational transformation of force constant of the $j\text{–}th$ normal mode $\left(k_j\right)_s = \xi^{-1}\left(k_j\right)_h$. Using this β-gravitational transformation for force constant in Eq. (53) evaluates β-gravitational transformation of vibrational reorganization energy,

$$(\lambda_i)_s = \xi^{-1}(\lambda_i)_h \tag{54}$$

Solvational reorganization energy for $\Delta e$ charge transferred between the reactants can be mathematically defined in terms of reactants ionic radii $a_1$ and $a_2$, their centre-to-centre separation distance $W$, refractive index "$n_{solv}$" and dielectric constants "$\varepsilon_{solv}$" of the solvent which at stronger gravitational field can be defined as;

$$(\lambda_o)_s = (\Delta e)^2\left(\frac{1}{2a_1} + \frac{1}{2a_2} - \frac{1}{W}\right)\left(\frac{1}{(n_{solv})^2_s} - \frac{1}{(\varepsilon_{solv})_s}\right) \tag{55}$$

β-gravitational transformation of mass will then certainly give β-gravitational transformation for dielectric constant i.e. $(\varepsilon_{solv})_s = \xi(\varepsilon_{solv})_h$ and β-gravitational transformation for refractive index of solvent i.e. $(n_{solv})_s = \xi^{-1/2}(n_{solv})_h$. Substituting gravitational transformation of refractive index and dielectric constant in Eq. (55) gives,

$$(\lambda_o)_s = \xi^{-1}(\lambda_o)_h \tag{56}$$

Therefore, from Eq. (54) and (56) gravitational transformation for total reorganization energy can be formulated as $\lambda_s = \xi^{-1}\lambda_h$. Thus, β-gravitational transformations for free energy, reorganization energy, collision frequency and ideal gas constant in Eq. (52) yields,

$$(k_{AB})_s = \xi^{-1}(k_{AB})_h \tag{57}$$

Now consider an electron transfer reaction in different perspective. Let $A, B$ are reactants and $X^*, X$ are hypothetical initial and final thermodynamic states of the system defined as intermediates.

$$A + B \rightleftharpoons X^* \rightleftharpoons X \longrightarrow P \qquad (58)$$

When reactants approach each other to undergo electron-transfer reaction suitable solvent fluctuation leads to formation of state $X^*$, which have the same atomic configuration of the reacting pair and of the solvent as that of the activated complex, while it owns electronic configuration of the reactant. $X^*$ have two choices either to form the reactant following disorganization of some of the oriented solvent molecules, or it transform to state $X$ by undergoing an electronic transition, $X$ possess the same atomic configuration as that of $X^*$ but it owns the electronic configuration of the products. The pair of states $X^*$ and $X$ together constitutes activated complex. Now considering that chemical system carrying out such electron transfer reaction is drifted towards source of gravity. At stronger gravitational field rate constant for electron transfer reaction expressed in terms of electronic coupling $(H_{AB})_s$ between the reaction intermediates of the electron transfer reaction (i.e., the overlap of the electronic wave functions of the two states) can be formulated as [37],

$$(k_{et})_s = \frac{2\pi}{\hbar}(H_{AB})_s^2 \frac{1}{\sqrt{4\pi\lambda_s R_s T}} exp\left(\frac{-((\triangle G^o)_s+\lambda_s)^2}{4\lambda_s R_s T}\right) \qquad (59)$$

β-gravitational transformations for free energy i.e., $(G^o)_s = \xi^{-1}\,(G^o)_h$ , electronic coupling $(H_{AB})_s = \xi^{-1}(H_{AB})_h$ , universal gas constant $R_s = \xi^{-1}R_h$ and total reorganization energy $\lambda_s = \xi^{-1}\lambda_h$ results in following gravitational transformation for electron transfer rate constant,

$$(k_{et})_s = \xi^{-1}(k_{et})_h \qquad (60)$$

### 5.4 Gravitational transformation of RRKM rate constant

If a system carrying out unimolecular reaction is drifted towards source of gravity then according to RRKM theory whcih unifies statistical RRK theory with transition state theory at high pressure limit will give following expression for the rate constant at stronger gravittaional field [30],

$$(k_{uni}^{\infty})_s = \frac{(k_B)_s T}{h}\frac{Q_r^{\dagger} Q_v^{\dagger} \exp\left(-(E_0)_s/(k_B)_s T\right)}{Q_r Q_v} \qquad (61)$$

β-gravitational transformations of Boltzmann constant and activation energy will also give following β-gravitational transformation for unimolecular rate constant,

$$(k_{uni}^{\infty})_s = \xi^{-1}(k_{uni}^{\infty})_h \qquad (62)$$

### 5.5 Gravitational transformation of rate of reaction

From basic knowledge of chemical kinetics, it is well known that the rate of a chemical reaction is defined as the rate of change of concentration "$C$" with respect to time $t$ [31, 32]. In case of gas phase reaction "$C$"is replaced by pressure "$P$"and number of molecules or atoms "$N$" in solid phase reactions (nuclear reactions). Now considering a chemical system with concentration of

"$C$" is towards source of gravity. Then at low gravitational field rate equation will be formulated as,

$$(r_n)_s = \frac{d[C]}{dt_s} = (k_n)_s[C]^n \tag{63}$$

Substituting β-gravitational transformation of rate constant in Eq. (63) gives β-gravitational transformation of rate of chemical reaction i.e.

$$(r_n)_s = \xi^{-1}(r_n)_h \tag{64}$$

Thus from Eq. (64) rate of reaction at stronger gravitational field i.e. near the suraface of planet is slower than at weaker gravitational field i.e . at appreciable height from the suraface of planet.

## 5.6 Gravitational transformation of Half Life

If a chemical system having initial concentration of reactants $C_o$ is drifted towards center of gravity. Then half-life period during which one half of the initial concentration $C_o$ of a reactant converts into products at stronger gravitational field can be defined as,

$$\left(t_{1/2}\right)_s = \frac{J}{(k_n)_s[C_o]^{(n-1)}} \tag{65}$$

$J$ is the coefficient for $n^{th}$ order reaction. Placing β-gravitational transformation of rate constant i.e. $(k_n)_s = \xi^{-1}(k_n)_h$ in Eq. (65) simply gives gravitational half- life equation,

$$\left(t_{1/2}\right)_s = \xi\left(t_{1/2}\right)_h \tag{66}$$

Gravitational transformation for half-life is completely the mirror image of Einstein's gravitational time dilation equation. This can explain time dilation at molecular level as Einstein's equation does in the physical world.

## 5.7 Relativistic Equilibrium Constant

Now considering a chemical reaction at chemical equilibrium [25],

$$aA + bB \rightleftharpoons cC + dD \tag{67}$$

Let the system where this chemical reaction at chemical equilibrium is carried out is drifted towards surface of heavy planet then equilibrium constant for this reaction can be expressed in terms of partition function at stronger gravitational field as,

$$\left(K_{eq}\right)_s = \frac{[Q_C]^c[Q_D]^d}{[Q_A]^a[Q_B]^b} exp\left(\frac{-\langle\triangle\varepsilon_o\rangle_s}{R_sT}\right) \tag{68}$$

Since difference in zero-point energies of reactants and products "$\langle\triangle\ \varepsilon_o\rangle_l$"and ideal gas constant undergoes β-gravitational transformations thus both of them together inhibits any change in magnitude of chemical equilibrium constant in the entire gravitational field i.e.

$$\left(K_{eq}\right)_s = \left(K_{eq}\right)_h = K_{eq} \tag{69}$$

System at chemical equilibrium should appear the same irrespective of its position in the gravitational field. Since chemical equilibrium constant is also defined as ratio of rate constants of forward and backward reactions of chemical system at equilibrium i.e. [25]

$$K_{eq} = (k_f)_s/(k_b)_s \tag{70}$$

Substituting β-gravitational transformations of forward and backward rate constants of chemical reactions making up chemical equilibrium in Eq. (70) will also keep chemical equilibrium constant in whole gravitational field. Thus, the amount of reactant and product in chemical equilibrium with one another remain the same independent of position of chemical system in the gravitational field.

# 6 Results and discussion

## 6.1 Discussion on gravitational statistical thermodynamics

The present theory shows that the Boltzmann constant is consistent with the gravitational transformation of energy spacing between permitted quantum levels of molecules. Since molecular transition between two states is dictated by the spacing among them which results in electronic transitions to be quicker than vibronic transitions and vibronic transitions to be quicker than rotational transitions. As general theory of relativity works on the principle that time flows slower near the surface of heavy planet i.e. at stronger gravitational field result in slower de-excitation of an excited state. This can only be afforded at the expense of decrease in energy spacing between different quantum states i.e. β-gravitational transformation of energy spacing between quantum levels i.e. $(\Delta\varepsilon)_s = \xi^{-1}(\Delta\varepsilon)_h$. Experimentally it is known that in molecules possessing lighter isotopes of elements constituting them have elevated quantum levels than those with heavier isotopes [38]. Thus, based on this observation, the gravitational mass dilation fully supports the gravitational transformation of energy spacing between different quantum levels of the system under study. In all three kind of statistics i.e. Fermi-Dirac statistics, Bose-Einstein statistics and Maxwell Boltzmann statistics describing distribution of the total number of particles among different quantum states in different scenarios there exists a common exponential factor i.e., $w_s = exp(-(\triangle\,\varepsilon)_s/(k_B)_sT)$ [25-26]. β-gravitational transformations of energy levels and Boltzmann constant together keep this exponential factor constant in the entire gravitational field. So, partition function of all three kinds of statistics and their distribution in the respected permitted energy states do not change with change in the strength of gravitational field due to this exponential factor. This will result in the number of molecules in particular energy levels to remain the same irrespective of the position of the system in the entire gravitational field. Thus, the ratio of the distribution of the total energy among the total number of molecules comprising the system under study and all of their internal degrees of freedom (translational, vibrational, rotational and electronic) will remain the same irrespective of the position of the system in the gravitational field. At stronger gravitational potential mass of molecules increases that will result in $\boldsymbol{\alpha}$-gravitational transformation i.e. $I_s = \xi I_h$. To conserve the law of conservation of angular momentum angular velocity decreases at stronger gravitational potential i.e., $\omega_s = \xi^{-1}\omega_h$. This will result in decrease of rotational speed of molecules at stronger gravitational field which supports the fact that time flows slowly near the source of gravity. The $\boldsymbol{\alpha}$-gravitational transformation of rotational inertia of molecules dictates rotational constant to have β-gravitational transformation at stronger gravitational potential i.e., $B_s = \xi^{-1}B_h$. This again supports the gravitational time dilation. The β-gravitational transformations of rotational constant and Boltzmann constant keep the rotational temperature same at all positions in the gravitational field. This makes rotational partition function to remain the same at all positions in the entire gravitational field. Since molecules execute vibrations, the $\boldsymbol{\alpha}$-gravitational transformation of the mass of atoms at stronger gravitational field will decrease the vibrational frequencies of all vibrational modes present in the molecules resulting in the β-gravitational transformation of vibrational frequency i.e., $\upsilon_s = \xi^{-1}\upsilon_h$, this again supports the slow passage of time near the source of gravity. As at surface of heavy planet time flows slowly as compared to weaker gravitational potential, so atoms in molecules will undergo slower compressions and extensions at stronger gravitational field. β-gravitational transformations of vibrational frequency and Boltzmann constant at stronger gravitational field keeps the vibrational temperature constant at all positions in the entire gravitational field. Similarly, this makes vibrational partition function to remain the same at all positions in the whole gravitational field. Electronic temperature and electronic partition function will not change with change in the strength of the gravitational field due to $\boldsymbol{\alpha}$-gravitational transformation of mass and β-gravitational transformations of velocities of electrons and their permitted quantum states. Thus, all kinds of molecular partition functions i.e., translational, rotational, vibrational, and electronic will remain same throughout the entire gravitational potential. Thus, $\boldsymbol{\alpha}$-gravitational transformation of mass of molecules and β-gravitational transformation of spacing between quantum levels at stronger gravitational potential keeps the translational, vibrational, rotational, and electronic partition functions constant at all gravitational potentials i.e. stronger and weaker potentials. Since translational, rotational, vibrational, and electronic temperature had already been mathematically proved invariant under Lorentz boosts [17]. Similarly, there also exists no gravitational transformation for translational, rotational, vibrational, and electronic temperature. Since all thermodynamic state variables are mathematically expressed in terms of partition functions together with Boltzmann constant. The β-gravitational transformation of Boltzmann constant will make all thermodynamic state functions to decrease near the source of gravity i.e. at stronger gravitational potential. At stronger gravitational potential mass of particles increases that result in decrease of all forms of their internal motions and thus, all molecular transformations defining thermodynamic state functions decreases at stronger gravitational potential which is in accordance with the slower flow of time near the source of gravity.

### 6.2 Discussion on gravitational chemical kinetics

Transition state theory is the most successful and universal theory of chemical kinetics for evaluating reaction rates [39]. This theory first time coins the concept of an activated complex called transition state which is responsible for conversion of reactants into products by making a translational sweep over the reaction coordinate. According $\boldsymbol{\alpha}$-gravitational transformation of mass transition state mass will increase i.e. $m_s^{\dagger} = \xi m_h^{\dagger}$ near source of gravity i.e. at stronger gravitational field. When a chemical system is drifted towards source of gravity then according to the β-gravitational transformation of velocity transition state velocity will decrease at stronger gravitational field i.e., $v_s^{\dagger} = \xi^{-1} v_h^{\dagger}$ so this will make the momentum of the transition state to remain same at all positions in the gravitational field i.e., $\Delta p_s = \Delta p_h = \Delta p$. Since de Broglie relation shows that mass and de Broglie wavelength associated with a transition state are inversely related to one another i.e., $\lambda_s = h/\triangle\, p_s$ [31]. As the momentum of transition state remain same at all positions in the entire gravitational field, so de Broglie associated with transition state remains same in the entire gravitational potential i.e., $\lambda_s = \lambda_h = \lambda$. This de Broglie wave associated with transition state should not be confused with de Broglie wave explained earlier, while formulating gravitational mass dilation. In that thought experiment stream of atoms or electrons are pumped out straight in a beam against gravitational potential leading to gravitational red-shift in their de Broglie wavelength, while here transition states for two same reactions are discussed which are carried out at different positions in a gravitational field, whose de Broglie wave remains same in entire gravitational field. At the stronger gravitational field transition state produced will be more massive than transition state produced at weaker gravitational potential, so to have same de Broglie wavelength transition state at different gravitational fields; it should possess smaller velocity at the stronger gravitational field i.e. near the source of gravity. This fully supports the idea that time flow slower near the source of gravity. According to Heisenberg's Uncertainty relation reaction coordinate can only be sufficient to accommodate transition state if it is of size equal to de Broglie wavelength i.e., $(\Delta q)_s \geq \lambda_s/2\pi$ [17]. Since de Broglie wavelength of transition state is same at all positions in the gravitational field i.e., this keeps reaction coordinate to be same at all positions in the gravitational field i.e., $(\Delta q)_s = (\Delta q)_h = \Delta q$. Accommodation of activated complexes along the reaction coordinate is mathematically defined as $\delta_s^{\dagger} = h/\sqrt{\left(2\pi m_s^{\dagger} (k_B)_s T\right)}$ it can be shown that it remains the same at all positions in the gravitational field by β-gravitational transformation of Boltzmann constant and **α**-gravitational transformation mass i.e., $\delta_s^{\dagger} = \delta_h^{\dagger} = \delta^{\dagger}$. At the stronger gravitational field, the average rate of passage of activated complexes over the barrier along the coordinate can be formulated mathematically as $r_s^{\dagger} = \sqrt{(k_B)_s T/2\pi m_s^{\dagger}}$ which on substituting β-gravitational transformation of Boltzmann constant and mass gives $r_s^{\dagger} = \xi^{-1} r_h^{\dagger}$. Therefore, at stronger gravitational field transition state will have less kinetic energy and thus, it will undergo slower translations over the reaction coordinate. β-gravitational transformation of Boltzmann constant equips the transition state with less kinetic energy and slow down its motion along the reaction coordinate. The rate constant of fastest reaction possible at the stronger gravitational field will be an order of $(k_B)_s T/h$ [40]. At the stronger gravitational field β-gravitational transformation of Boltzmann constant will decrease the rate constant by decreasing the frequency of the passage of reactants through the transition state. β-gravitational transformation of rate constant holds for all kinds of reactions regardless of what type of kinetics they follow i.e., either zero order, first order, second order or third order etc. Transitions state theory defines the Arrhenius factor in terms of thermodynamic state functions. Since for all thermodynamic state functions and universal gas constant undergoes β-gravitational transformation thus dictating Arrhenius factor to have the same transformation. Arrhenius factor appearing as a pre-exponential factor in the rate equation controls the frequency for the passage through the transition state. **α**-gravitational transformation of a mass of reacting atoms and molecules decreases the Arrhenius at stronger gravitational potential which fully supports the slower passage of time near the source of gravity or surface of heavy planet. Rate constant gives a quantitative picture of the speed of reaction; as the rate constant decreases, the rate of reaction also decreases, at the stronger gravitational potential i.e. near the source of gravity. This is for the first time mathematically shown in current work that the rate of reaction has a gravitational transformation. Gravitational transformation of the half-life period equation as shown in Eq. (66) again supports Einstein's gravitational time dilation. This shows a deep connection between the present theories of rates of reactions with the general theory of relativity.

Collision theory majorly focusing on the kinetics of bimolecular reactions expresses the rate of reactions as the frequency of bimolecular collisions. Rate of reaction is dictated by the number of fruitful collisions occurring per second. Frequency of collision is responsible for the rate of reaction. Rate constant in collision theory is a product of the relative velocity of colliding molecules and collision cross-section area undergoing in the bimolecular collision. $\boldsymbol{\alpha}$-gravitational transformation of a mass of molecules decreases the relative velocities between the colliding molecules. β-gravitational transformation of Boltzmann constant also demotes the energy available per molecule, Thus, dropping up the relative velocity between the molecules. Collision frequency defines the rate of reaction as a product of the area of collision cross-section, mole densities and relative velocity of colliding molecules. Since collision cross-section area and mole densities (concentration) remains the same at all positions in the entire gravitational field, while relative velocity decreases that give β-gravitational transformation of collision frequency as shown in Eq. (51). Decrease in collision frequency of molecules at stronger gravitational field slows down the reaction rate. β-gravitational transformation of collision frequency transformation fully agrees with the idea of slowing down of time near the source of gravity given by general theory of relativity. Since time flows slower at stronger gravitational potential, so does collision frequency of molecules also decrease near the source of gravity.

The Marcus theory is a statistical mechanical approach employing potential energy surfaces to describe several important redox processes in chemistry and biology [36,37]. These redox reactions occur on a scale much faster than the nuclear vibrations. Therefore, the nuclei do not appreciably move during electron transfer phenomenon. During the transfer energies of the donor and acceptor, orbitals must match. The energy levels of the donor and acceptor orbitals in the reactants and products come in continual flux due to internal nuclear and the solvent motions. For a successful transfer, the donor and acceptor molecules must attain definite geometries and suitable solvation arrangements that result in the matching of energy levels between the donor and acceptor orbitals. The nuclei of donor and acceptor molecules relax to their optimum positions once electron transfer has occurred. The energy needed to modify the solvation sphere and internal structures making the donor and acceptor orbital of the same energy is defined as reorganization energy. This energy is a barrier to electron transfer and appears in two forms. One of them is vibrational reorganization energy $(\lambda_i)_s$ that is the energy difference between changes in bond length, angles etc. which occur upon electron transfer. $\boldsymbol{\alpha}$-gravitational transformation of a mass of nuclei of reacting species decreases their characteristic oscillation frequency and thus, reduces the force constant at a stronger gravitational field. So, energy needed to bring changes in bond length, angles etc. For a successful transfer of electron rate at the stronger gravitational field will decrease as compared to same reaction conducted at weaker gravitational potential as shown in Eq. (54). While solvational reorganization energy $(\lambda_o)_s$ that counts the energy required in the reorganization of the solvent shell is shown in Eq. (55). When electron transfer reaction is carried out at stronger gravitational potential $\boldsymbol{\alpha}$-gravitational transformation of mass makes density to increase. Therefore, at stronger gravitational field solvent becomes denser than at weaker gravitational potential resulting in the $\boldsymbol{\alpha}$-gravitational transformations of refractive index and dielectric constant making solvational reorganization energy to have β-gravitational transformation for the successful transfer of electron which is shown in Eq. (56). β-gravitational transformations of solvational reorganization energy and vibrational reorganization energy together makes β-gravitational transformation of total reorganization energy possible. This is similar to gravitational transformations of other thermodynamic state functions expressed earlier in this work. Electron transfer reaction occurs via a very slight spatial overlap of the electronic orbitals of the two reacting molecules in the activated complex. The assumption of slight overlap is defined in terms of electronic coupling between intermediates $X^*$ and $X$. The intermediate state $X^*$ has two choices either to disappear and reform reactants or undergo electronic jump to form a state $X$ in which the ions resemble characteristics of products. Rate constant equation for such electron transfer reaction have electronic coupling $(H_{AB})_s$ as a pre-exponential factor that defines the electronic overlap between the intermediates $X^*$ and $X$. According to Heisenberg Uncertainty principle, *the greater the overlap shorter will be the lifetimes of states $X^*$ and $X$* [35]. At stronger gravitational field $\boldsymbol{\alpha}$-gravitational transformation of mass transformation will decrease the velocity of electrons keeping the net momentum same while space occupied by these electrons remains unchanged due to momentum-space Uncertainty principle i.e., bond length remains same at all positions in the gravitational field. This decrease in electronic oscillations confined in the same space (orbitals) will made the spatial overlap of orbitals to decrease at the stronger gravitational field, and thus β-gravitational transformation of electronic coupling is obtained as $(H_{AB})_s = \xi^{-1}(H_{AB})_h$. Thus, rate constant for electron transfer reaction decreases at stronger gravitational field supports slow passage of time near the source of gravity.

# 7 Application

### 7.1 Gas phase reaction of hydroxyl radical with 2-methyl-1-propanol

To demonstrate the effectiveness of the present theory, we consider example of gas-phase reactions of hydroxyl radical with 2-methyl-1-propanol [41] conducted on the surface of earth. Rate constant of reaction has been experimentally measured on the surface of earth and has been reported [41]. Now consider same reaction is carried out in stationary laboratory named "$CZ$" build at a distance "$h$" equal to 939 km away from the event horizon of Black hole named as "$M87$" [42]. Laboratory "$CZ$"should be kept stationary to avoid emergence of relativistic effects from special relativity. For observing only effect of gravity on rate constant of chemical reaction under study temperature and pressure i.e., 298 K and 1 atm should be same at the surface of earth and laboratory "$CZ$". Now laboratory at the surface of earth is at weaker gravitational field as compared to laboratory "$CZ$" which is at stronger gravitational filed. This is because even at distance of 939 km gravity of Black hole "$M87$" will be much stronger than gravity at surface of earth [42]. As it has been mentioned above that physical quantities whose magnitude is affected by strength of gravitational field will have their exact value in free space in absence of gravity. Earth's surface will be at a stronger gravitational field as compared to a laboratory in free space where there is no gravity. From Eq. (43) mathematical relation giving comparison of exact rate constant of chemical reaction "$k$" in free space in absence of any gravitational filed as compared to rate constant of reaction "$k_{earth}$" on the surface of earth can be formulated as;

$$k = k_{earth}/\sqrt{1 - 2GM_{earth}/R_{earth}c^2} \quad (71)$$

Laboratory "$CZ$" at distance of 939 km away from the event horizon of Black hole "$M87$" is at stronger gravitational field as compared to laboratory in free space where there is no gravity. From Eq. (43) mathematical relation giving comparison of exact rate constant of chemical reaction in free space in absence of any gravitational filed as compared to rate constant of reaction "$k_{M87}$" conducted in laboratory "$CZ$" be formulated as;

$$k = k_{M87}/\sqrt{1 - 2GM_{M87}/(R_{M87} + h)c^2} \quad (72)$$

Comparing Eq. (71) and (72) will give following relation;

$$k_{M87} = k_{earth}\left(\sqrt{1 - 2GM_{M87}/(R_{M87} + h)c^2}/\sqrt{1 - 2GM_{earth}/R_{earth}c^2}\right) \quad (73)$$

Placing the values universal gravitational constant "$G$"; mass of earth "$M_{earth}$"; mass of Black hole "$M_{M87}$"; radius of $R_{earth}$ and Schwarzschild radius of black hole "$R_{M87}$"; distance "$h$" of laboratory "$CZ$" from event horizon of "$M87$" and rate of chemical reaction reported on the surface of earth "$r_{earth}$" in Eq. (73) will give;

$$k_{M87} = k_{earth}/10 \quad (74)$$

From Eq. (74) it can be inferred that in laboratory "$CZ$" at a distance of 939 km away from the event horizon of black hole "$M87$" rate constant of chemical reaction will be reduced 10 times in magnitude than rate constant of same reaction on the surface of earth. Provided temperature and pressure are same at surface of earth and inside laboratory "$CZ$" where reaction is being conducted. Effect of gravity "$M87$" on rate constant of this gas phase reaction conducted in laboratory "$CZ$" and surface of earth is elaborated in Table 1.

**Table 1.** Comparison of rate constants at surface of earth's gravitational field and in laboratory "$CZ$" build at a distance of 939 km from event horizon of Black hole M87.

| Rate constant at surface of earth[41] ($cm^3 molecule^{-1} s^{-1}$) | Rate constant of reaction in "$CZ$" ($cm^3 molecule^{-1} s^{-1}$) |
|---|---|
| $0.92\times10^{-11}$ | $0.92\times10^{-12}$[a] |
| [a] This value has been developed from Eq. (74) | |

## 7.2 Rearrangement of *syn* and *anti*-aldehyde conformers to oxazole and ketene

To further elaborate the use of current theory let consider the rearrangement of two conformers of aldehyde to oxazole and ketene. The *syn* and *anti*-aldehyde conformers of triplet 2-formyl-3-fluorophenylnitrene generated in a nitrogen matrix by UV-irradiation of azide precursor spontaneously rearranges to the corresponding 2,1-benzisoxazole and imino-ketene respectively [43]. Rate constants for these transformations reported are $1 \times 10^{-3}$ $s^{-1}$ and $6 \times 10^{-3}$ $s^{-1}$ at 10 and 20 K [43]. If same reactions are carried at the height of 20200 km from the surface of earth keeping temperature constant as that was on the surface of earth i.e., 10 and 20 K respectively then applying again Eq. (43) will give rate constants to be $1.000000017 \times 10^{-3}$ $s^{-1}$ and $2.000000101 \times 10^{-3}$ $s^{-1}$. Thus, the rate of reaction has increased at weaker gravitational field i.e., 20200 km from the surface of earth. To study solely the effect of gravitational field on rate of chemical reaction all the external parameters that can affect the reaction rate i.e., temperature, volume and pressure etc. should be kept same at two gravitational potentials where reactions must be carried out. Like in present case theoretically temperature for both the reactions at the surface of earth (stronger gravitational field) and at height of 20200 km from the surface of earth (weaker gravitational field) are kept same i.e., 10 and 20 K respectively.

## 8 Conclusions

All gravitational transformations developed in current theory fully support the slower flow of time near the surface of heavy planets, which is one of the main experimentally verified facts of general relativity. The slower decay of excited states near the source of gravity is achieved by squeezing the spacing between permitted quantum levels. This result has been proved mathematically to obtain the β-gravitational transformation of energy spacing. Thus, gravity seems to kill quantization. The behavior of gravitational force appears to be quantumphobic, which tries to reduce the width of quantization. For heavy planets like Earth, this effect will not be large enough, but massive bodies like black holes will have a pronounced effect on quantization. Gravity had already been reported to induce decoherence in micro-scale quantum systems and therefore account for the emergence of classicality [44]. The gravitational transformations of all thermodynamic state functions formulated in this paper are completely compatible with gravitational time dilation. The rate constant of reaction decreases near a stronger gravitational potential, which thus successfully explains the slowing down of the rate of all molecular processes near the source of gravity. Thus, slower time flow in molecular rate processes at stronger gravitational potential is a consequence of an increase in the mass of subatomic particles constituting atoms and molecules and a decrease in the energy spacing of all quantum levels associated with them.

**Declaration of interest:** Author declares no competing conflict of interests.

**Author contributions:** Author have accepted responsibility for the entire content of this manuscript and approved its submission.

**Funding Information:** No funding was recieved for this work.

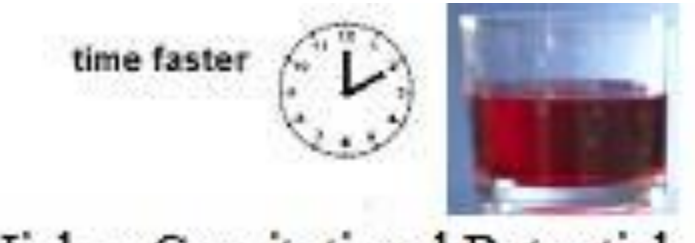

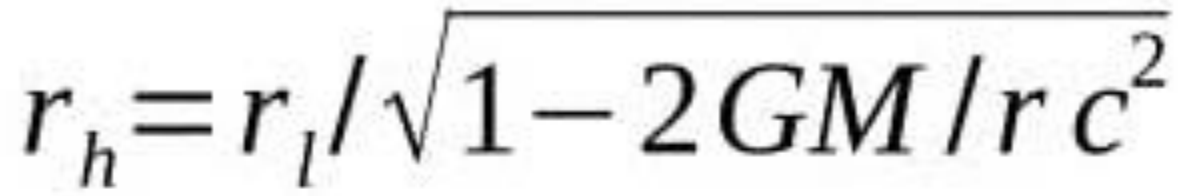

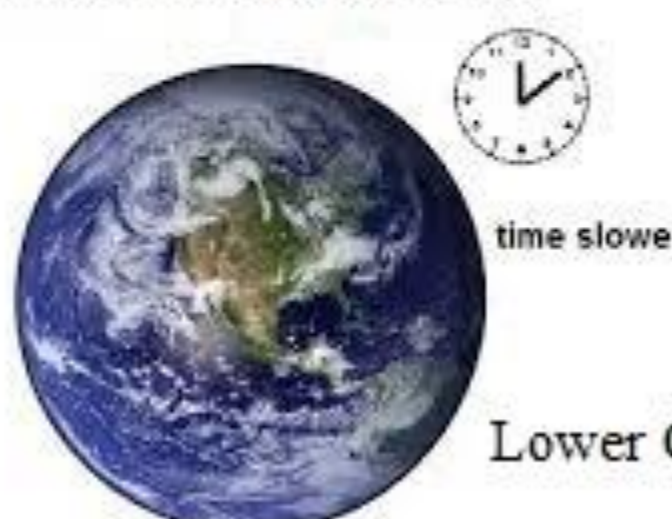

Rate of Chemical Reactions slows down at Lower Gravitational potential.

Lower Gravitational Potential

- Stronger gravitational field is lower gravitational potential
- Weaker gravitational field is higher gravitational potential
- In general theory of relativity lower and higher refers to the distance from source of gravity